\documentstyle[12pt]{article}
\title{{\bf  Gluon penguin enhancements to inclusive charmless decays
of b quark in the 2HDM with flavor changing couplings }}
\author{ Zhenjun Xiao$^{(1,2)}$
\thanks{E-mail: zxiao@ibm320h.phy.pku.edu.cn},
Chong Sheng Li$^{(1)}$, Kuang-Ta Chao$^{(1)}$ \\
{\small 1. Department of Physics,
Peking University, Beijing, 100871 P.R. China.} \\
{\small 2. Department of Physics, Henan Normal University, Xinxiang, 453002
P.R. China.}  \\ }

\begin{document}
\maketitle

\begin{abstract}
We calculate the enhancements to the inclusive charmless decays of b quark,
$b \to s g, s q\bar{q}, sgg$, from gluon penguin diagrams induced by the
charged and neutral Higgs bosons $(H^{\pm}, h^0, H^0$, and $A^0)$ in the
Two-Higgs-Doublet Model with flavor-changing couplings. Within the
considered parameter space, the new contributions from charged Higgs boson
are dominant. After including the new contributions, the branching ratio
$BR(b \to sg)$ ($q^2 =0$) can be increased form $\sim 0.2\%$ in the
standard model to $4.4\%$ and $2.6\%$ in the two-Higgs-doublet
model for $m_{H^+}=100$ and $200$ GeV, respectively. The new
contribution to the decay mode $b \to s q \bar{q}$ with $q=(u,d,s)$
is, however, numerically small and peaked at the lower $q^2$ region.
The new contribution to $b \to s gg$ can also be neglected.
\end{abstract}
\vspace{.5cm}


\noindent
PACS numbers: 13.25.Hw, 12.15.Ji, 12.38.Bx, 12.60.Fr

\noindent
Key words:  gluon penguin, charmless decay, b quark, charged Higgs

\newcommand{\beq}{\begin{eqnarray}}
\newcommand{\eeq}{\end{eqnarray}}

\newcommand{\bsg}{b \to s g}
\newcommand{\bsgs}{b \to s g^*}
\newcommand{\bsqq}{b \to s q \bar{q}}
\newcommand{\bsuu}{b \to s u \bar{u}}
\newcommand{\bsss}{b \to s s \bar{s}}
\newcommand{\bsgg}{b \to s g g}

\newcommand{\brbsg}{BR(b \to s g) }
\newcommand{\brbsuu}{BR(b \to s u \bar{u}) }
\newcommand{\brbsqq}{BR( b \to s q \bar{q})}
\newcommand{\brbsgg}{BR( b \to s g g)}
\newcommand{\brbsgs}{BR( b \to s g^*)}

\newcommand{\mhp}{M_{H^{+}}}
\newcommand{\mhz}{M_{h^0}}
\newcommand{\mhb}{M_{H^0}}
\newcommand{\mha}{M_{A^0}}

\newcommand{\ati}{\tilde{A}_i}
\newcommand{\bti}{\tilde{B}_i}
\newcommand{\cti}{\tilde{C}_i}
\newcommand{\dti}{\tilde{D}_i}

\newcommand{\att}{\tilde{A}_t}
\newcommand{\btt}{\tilde{B}_t}
\newcommand{\ctt}{\tilde{C}_t}
\newcommand{\dtt}{\tilde{D}_t}

\newcommand{\atu}{\tilde{A}_u}
\newcommand{\btu}{\tilde{B}_u}
\newcommand{\ctu}{\tilde{C}_u}
\newcommand{\dtu}{\tilde{D}_u}

\newcommand{\atc}{\tilde{A}_c}
\newcommand{\btc}{\tilde{B}_c}
\newcommand{\ctc}{\tilde{C}_c}
\newcommand{\dtc}{\tilde{D}_c}

\newpage

Among radiative decays of B meson, $b \to s g$ is theoretically
clean, phenomenologically interesting
and sensitive to new physics beyond the standard model (SM),
for example, the two-Higgs-doublet model (2HDM)
\cite{2hdm}, the minimal supersymmetric standard model(MSSM)
\cite{misiak97,abel98} and Technicolor models\cite{xiao99}.

In the SM, $\brbsg \sim 0.2\%$ for on-shell gluon
and $BR(B \to X_{no charm}) \sim 1 -2\%$\cite{ciuchini94}.
According to the studies in ref.\cite{kagan98}, an enhanced
$b \to sg$ is favored phenomenologically since such enhancement
is very helpful for example to decrease the averaged charm
multiplicity $n_c$ and the semileptonic branching ratio\cite{hou91}
and to increase the kaon yields\cite{kagan96}. For the large
$BR(B \to \eta' X_s)$ measured by CLEO\cite{cleo98}, one can also
give a plausible interpretation from an enhanced
$\bsg$\cite{kagan98,hou98}. The contributions to the ratio $\bsg$
in 2HDM without flavor changing(FC) couplings were calculated in
refs.\cite{hou94,kagan95}, and the authors found that this ratio less than
$0.7\%$ for $m_H \geq 200$ GeV and $\tan{\beta}\sim 5$ in both Type I and II
2HDM \cite{2hdm}. Such a enhancement is not large enough to meet the
requirement. The possibility of a large rate for $\bsg$ in
supersymmetric models were also studied in ref.\cite{bertolini87}.

In this letter, we calculate the contributions to the inclusive charmless
decays $\bsg$ ($q^2=0$), $\bsqq$ and $\bsgg$ from the gluon penguin
diagrams induced by the exchange of charged and neutral Higgs bosons
in the so-called Model III: the two-Higgs-doublet model with FC
couplings\cite{hou92,atwood97}. We found that  the branching ratio
$\brbsg$ can be increased from $\sim 0.2 \%$ in the SM to $ 2-4\%$ level
in the Model III.  So large enhancement is still consistent with the  CLEO
limit: $BR(b \to s g^*) < 6.8\%$ at $90\% C.L.$\cite{cleo98b},
and will be very helpful to resolve the experimental puzzles\cite{kagan98}.

In the 2HDM, the tree level flavor changing
scalar currents(FCSC's) are absent if one introduces an {\it ad
hoc} discrete symmetry to constrain the 2HDM scalar potential and
Yukawa Lagrangian. Lets consider a Yukawa Lagrangian
of the form\cite{atwood97}
\beq
{\cal L}_Y &=&
\eta^U_{ij}\bar{Q}_{i,L} \tilde{\phi_1}U_{j,R} +
\eta^D_{ij}\bar{Q}_{i,L} \phi_1 D_{j,R}
+\xi^U_{ij}\bar{Q}_{i,L} \tilde{\phi_2}U_{j,R}
+\xi^D_{ij}\bar{Q}_{i,L} \phi_2 D_{j,R}+ h.c., \label{leff}
\eeq
where $\phi_{i}$ ($i=1,2$) are the two Higgs doublets,
 $\tilde{\phi}_{1,2}=i\tau_2 \phi^*_{1,2}$, $Q_{i,L}$  with
 $i=(1,2,3)$ are the left-handed quarks,
$U_{j,R}$ and $D_{j,R}$  are the right-handed  up- and down-type
quarks,
while $\eta^{U,D}_{i,j}$  and $\xi^{U,D}_{i,j}$ ($i,j=1,2,3$ are
family index ) are generally the nondiagonal matrices of the  Yukawa
coupling. By imposing the discrete symmetry $(\phi_1 \to - \phi_1,
\phi_2 \to \phi_2, D_i \to - D_i, U_i \to  \mp U_i) $ one obtains
the so called model I and II.

In this letter, we will consider the third type of 2HDM: the so
called Model III \cite{hou92}: no discrete symmetry is imposed
and both up- and down-type quarks then have FC couplings with
$\phi_1$ and $\phi_2$. In Model III, there are five physical
Higgs bosons: the charged
scalar $H^{\pm}$, the neutral CP even scalars $H^0$ and $h^0$
and the CP odd pseudoscalar $A^0$. After the rotation that
diagonalizes the mass matrix of the quark fields, the Yukawa
Lagrangian of quarks  are the form \cite{atwood97},
\beq
{\cal L}_Y^{III} =
\eta^U_{ij}\bar{Q}_{i,L} \tilde{\phi_1}U_{j,R} +
\eta^D_{ij}\bar{Q}_{i,L} \phi_1 D_{j,R}
+\hat{\xi}^U_{ij}\bar{Q}_{i,L} \tilde{\phi_2}U_{j,R}
+\hat{\xi}^D_{ij}\bar{Q}_{i,L} \phi_2 D_{j,R} + h.c.,
\label{lag3}
\eeq
where $\eta^{U,D}_{ij}=m_i\delta_{ij}/v$ correspond to the diagonal
mass matrices of quarks and $v\approx 246 GeV$ is the vacuum
expectation value of $\phi_1$, while the neutral and charged FC
couplings will be \cite{atwood97,sher87}
\beq
\hat{\xi}^{U,D}_{neutral}&=& \xi^{U,D}, \nonumber\\
\hat{\xi}^{U}_{charged}&=& \xi^{U}V_{CKM}, \ \
\hat{\xi}^{D}_{charged}= V_{CKM} \xi^{D}, \label{cxiud}
\eeq
where $V_{CKM}$ is  the Cabibbo-Kabayashi-Maskawa mixing matrix\cite{ckm},
and
\beq
\xi^{U,D}_{ij}=\frac{\sqrt{m_im_j}}{v} \lambda_{ij}.\label{xiij}
\eeq
In this letter we assume that $\lambda_{ij}$ are real because we
here do not consider the possible effects of CP violation induced
by the phase of $\lambda_{ij}$.

As pointed in ref.\cite{atwood97},  the experimental data
of $K^0-\bar{K}^0$ and $B_d^0-\bar{B}_d^0$ mixing processes
put severe constraint on  the FC couplings involving
the first generation of quarks\cite{atwood97}:
$ (\lambda_{ui}, \lambda_{dj}) \ll 1 $ for $i,j=(1,2,3)$.
We here will enforce the same constraint:
$\lambda_{ui, dj}=0$ for $i,j=(1,2,3)$. And we also assume
that $ \lambda_{ij} = 1$, for $i,j=(2,3)$.

Direct searches for Higgs bosons in 2HDM at LEP II\cite{lep2}
place the following mass imits: $\mhp >56 GeV$, $ \mhz > 77 GeV$,
$\mha > 78 GeV$. From the CLEO data of $BR(B \to X_s \gamma)$,
some constraint on $\mhp$ in Model III can also be derived
\cite{atwood96,chao99}. If one uses new CLEO result \cite{cleo99} of
$B(B \to X_s \gamma)\leq 4.77 \times 10^{-4}$ at the $3\sigma$
level, the constraint $\mhp \geq 400$ GeV can be read off directly
from the Fig.2 of ref.\cite{atwood96}. According to the studies in
ref.\cite{chao99}, the existence of a charged Higgs boson with
$\mhp \sim 200$ GeV is still allowed. In this letter, therefore,
we will consider the mass range of $100$ GeV to $800$ GeV for all
Higgs bosons in Model III.

As for additional constraint on the Higgs boson masses from
some other processes \cite{atwood97,chao99}, they are
not good enough to compete with the $F^0-\bar{F}^0$ mixing
processes.

In the SM, the magnetic-penguin induced $bsg$ coupling leads to
$b \to sg^*$ transitions where $g^*$ could be light-like ( $q^2=0$,
on-shell gluon), or time-like ($q^2 >0$, off-shell gluon). Under
the spectator approximation, there are basically three
types of subprocesses: $b \to sg$, $b \to s q \bar{q}$ for
$q=u,d,s$, and $b \to s gg$. It is
straightforward to find the effective $bsg$ coupling by explicit
calculation, or by making appropriate changes to the $sd\gamma $
vertex of ref.\cite{inami81},
\beq
\Gamma_{\mu}(q^2)= \frac{ig_s }{4\pi^2}
\bar{u}_s(p_s) T^a V_{\mu}(q^2) u_b(p) \label{eq:gammaa}
\eeq
with
\beq
V_{\mu}(q^2) &=&\frac{g_2^2}{8 M_W^2} \left \{
\left ( q^2g_{\mu \nu} -q_{\mu} q_{\nu}\right )\gamma^{\nu}
\left [ F_1^L(q^2)L + F_1^R(q^2)R \right ] \right. \nonumber\\
&& \left.  + i \sigma_{\mu \nu} q^{\nu}
\left [ m_s F_2^L(q^2)L + m_b F_2^R(q^2)R \right ] \right \},
 \label{eq:vmu}
\eeq
where $p$ ($p_s$) is the b (s) quark momentum, $q=p -p_s$ is the
gluon momentum, $F_1^{L,R}$ and $F_2^{L,R}$ are the electric and
magnetic form factors, $g_s$ ($g_2$) is the QCD (electroweak)
coupling constant,
$L,R=(1\mp \gamma_5)/2$ are the chirality projection operators
and $T^a$ with $a=1, \cdots , 8$ are the $SU(3)_C$ generators.

If the heavy top quark is the internal quark in the penguin diagram,
the terms proportional to $ m_b^2/M_W^2, m_s^2/M_W^2, q^2/m_t^2$ can
be neglected and we then have\cite{inami81,hou88}
\beq
F_1^L(0)&=& \sum_{i}v_i f_1(x_i), \ \ F_1^R(0)=0,
    \label{f1lr0}\\
F_2^L(0)&=& F_2^R(0)=\sum_{i}v_if_2(x_i),\label{f2lr0}
\eeq
with
\beq
f_1(x_t)&=&\frac{1}{12(1-x_t)^4} \left [ 18x_t -29x_t^2 +
10x_t^3 + x_t^4 -(8 -32x_t +18 x_t^2)\ln x_t \right],
\label{f1xt} \\
f_2(x_t)&=&\frac{-x_t}{4(1-x_t)^4}\left [ 2+ 3x_t-6x_t^2 + x_t^3
+6x_t\ln x_t \right], \label{f2xt}
\eeq
where $x_t=m_t^2/M_W^2$, $v_i =V_{is}^*V_{ib}$ for $i=u,c,t$.

For decay $b \to s g$, $ q^2_{max} \approx 20 GeV^2$ and the
assumption $q^2 \ll m_i^2$ which would justify the replacement of
the form factors with their values at $q^2=0$ is valid only for top
quark, but not for the light $u$ and $c$ quarks
\cite{abel98,atwood97b}. The correct $f_1(x_j)$
for $j=u,c$ have been given for example in ref.\cite{abel98}, and
we also calculated the $f_2(x_j)$ for $j=(u,c)$ by using the same
technique as refs.\cite{abel98,atwood97b},
\beq
f_1(x_j,q^2)&=&\frac{10}{9}-\frac{2}{3}\ln x_j + \frac{2}{3z_j}
-\frac{2(1+2z_j)}{z_j}g(z_j),\label{f1xuc}\\
f_2(x_j,q^2)&=& -x_j( \ln x_j + 2 g(z_j) ), \label{f2xuc}
\eeq
where $z_j=q^2/(4m_j^2)$, and the explicit expression of $g(z_j)$ can
be found for example in ref.\cite{abel98}.
For $q^2 > 4m_j^2$, the internal $u$ or $c$ quark are on mass-shell,
 an absorptive part therefore appears for $f_1(x_j)$ and
$f_2(x_j)$, respectively.

In the Model III, an effective $bsg$ coupling can  also be induced
by the gluon penguin diagram via  exchanges of charged and neutral
Higgs bosons $(H^0, h^0, A^0, H^{\pm})$ as depicted in
Fig.1. We will evaluate the penguin diagrams involving $H^{\pm}$
and then extend the calculation to the neutral Higgs bosons. We
will follow the same procedure as that in the SM. We will use
dimensional regularization to regulate all the ultraviolet
divergence in the virtual loop corrections and adopt the
$\overline{MS}$ renormalization scheme. The necessary Feynman
rules can be obtained from the Lagrangian in Eq.(\ref{lag3}).

The Feynman diagrams with light internal $u$ and $d$ quarks do not
contribute since we have assumed that $\lambda_{ui,dj} =0$ for $i,j
=(1,2,3)$.
By explicit analytical calculations, we can extract out the form factors
$F_1^{L,R}$ and $F_2^{L,R}$ which describe
the new contributions from the neutral and charged Higgs bosons.

For the case of charged Higgs boson $H^\pm$, the effective $bsg$ vertex
is of the form,
\beq
\Gamma_{\mu}(q^2) &=& \frac{ig_s}{4\pi^2}\frac{g_2^2}{8 M_W^2}
 \bar{u}_s(p_s) T^a \left \{
 \left ( q^2g_{\mu \nu} -q_{\mu} q_{\nu}\right )\gamma^{\nu}
\left [ \tilde{F}_1^L L + \tilde{F}_1^R R \right ]\right. \nonumber\\
&&\left.  + i \sigma_{\mu \nu} q^{\nu}
\left [ m_s \tilde{F}_2^L L + m_b \tilde{F}_2^R R \right ]\right \} u_b(p),
\label{gammac}
\eeq
where
\beq
\tilde{F}_{1}^L(x_i)
&=&\frac{1}{\mhp^2} \left [ C_1(x_i)-C_{11}(x_i)\right ] \bti,
 \label{f1la}\\
\tilde{F}_{1}^R(x_i)
&=&\frac{1}{\mhp^2}\left [ C_1(x_i)-C_{11}(x_i)\right ]\ati ,
 \label{f1ra}\\
\tilde{F}_{2}^L(x_i)  &=& \frac{1}{\mhp^2}\left \{
3 C_{11}(x_i)\left [ \frac{m_b}{m_s}\ati + \bti \right ]
-C_1(x_i) \left [ \frac{m_b}{m_s} \ati + \bti + \frac{2m_i}{m_s}
 \cti \right] \right \}\label{f2la}\\
\tilde{F}_{2}^R(x_i)&=& \frac{1}{\mhp^2}\left \{
3C_{11}(x_i)\left [ \frac{m_s}{m_b}\ati + \bti \right ]
-C_1(x_i) \left [ \frac{m_s}{m_b} \ati
+ \bti + \frac{2m_i}{m_b} \dti \right] \right \}
\label{f2ra}
\eeq
with
\beq
C_{1}(x_i)&=&\frac{3-x_i}{4(1-x_i)^2} + \frac{1 }{2(1-x_i)^3}
    \log x_i, \label{c1xi}\\
C_{11}(x_i)&=& \frac{11-7x_i + 2x_i^2}{36(1-x_i)^3}
    + \frac{1 }{6(1-x_i)^4} \log x_i, \label{c11xi} \\
\ati &=& \frac{v^2}{2} \left ( V_{CKM}\xi^D \right )^*_{is}
\left ( V_{CKM} \xi^D \right )_{ib}, \ \
\bti = \frac{v^2}{2} \left ( \xi^U V_{CKM} \right )^*_{is}
\left ( \xi^U V_{CKM} \right )_{ib},  \label{ati}\nonumber\\
\cti &=& -\frac{v^2}{2} \left ( V_{CKM}\xi^D \right )^*_{is}
\left ( \xi^U V_{CKM} \right )_{ib}, \ \
\dti = -\frac{v^2}{2} \left ( \xi^U V_{CKM} \right )^*_{is}
\left ( \xi^U V_{CKM} \right )_{ib}, \label{cti}
\eeq
where $\xi^{U,D}$ have been given in Eq.(\ref{xiij}),
$x_i=m_i^2/\mhp^2$ for $i=c,t$. The  expressions
of $C_1(x_i)$ and $C_{11}(x_i)$ function are exact for the case of heavy
internal top quark, and correct approximately ( the error is less than $5\%$)
for the case of internal c quark.

By the same procedure, we
get the effective $bsg$ vertex induced by the neutral Higgs bosons,
\beq
\Gamma_{\mu}^{N}(q^2)= \frac{ig_s }{4\pi^2} \frac{g_2^2}{8M_W^2}
\sum \bar{u}_s(p_s) T^a
\left [  F_1 \left ( q^2g_{\mu \nu} -q_{\mu} q_{\nu}\right)
\gamma^{\nu}+ i  F_2 \sigma_{\mu \nu} q^{\nu} \right ]u_b(p)
\label{gammahn}
\eeq
where the summation over all three kinds of neutral Higgs bosons is
understood.

For the CP even scalar Higgs boson $h^0$, we have
\beq
F_1^{h^0}&=&\frac{1}{4\mhz^2} \sum_{i=s,b}
\left [ 3C_{11}(x_i)-C_1(x_i) \right ]\nonumber\\
&&\cdot
\left [ \sqrt{m_s m_b}m_i \lambda_{is}\lambda_{ib}\cos^2{\alpha}
-\sqrt{m_s m_b}m_b \lambda_{bs}\sin \alpha \cos \alpha \right ],
\label{f1h0}\\
F_2^{h^0}&=& \frac{1}{4 \mhz^2} \sum_{i=s,b}
\left [ -m_b ( 3C_{11}(x_i)-C_1(x_i) )
- 2m_i C_1(x_i) \right ] \nonumber\\
&& \cdot
\left [ \sqrt{m_s m_b}m_i \lambda_{is}\lambda_{ib}\cos^2{\alpha}
-\sqrt{m_s m_b}m_b \lambda_{bs}\sin \alpha \cos \alpha \right ],
\label{f2h0}
\eeq
where $x_i=m_i^2/\mhz^2$.

For the CP even scalar Higgs boson $H^0$, we have
\beq
F_1^{H^0}&=&\frac{1}{4\mhb^2} \sum_{i=s,b}
\left [ 3C_{11}(x_i) -C_1(x_i) \right ]\nonumber\\
&& \cdot
\left [ \sqrt{m_s m_b}m_i \lambda_{is}\lambda_{ib}\sin^2{\alpha}
+\sqrt{m_s m_b}m_b \lambda_{bs}\sin{\alpha}\cos{\alpha} \right ],
\label{f1hb0}\\
F_2^{H^0}&=&\frac{1}{4 \mhb^2} \sum_{i=s,b}
\left [ -m_b ( 3C_{11}(x_i) -C_1(x_i) )
- 2m_i C_1(x_i) \right ] \nonumber\\
&& \cdot
\left [ \sqrt{m_s m_b}m_i \lambda_{is}\lambda_{ib}\sin^2{\alpha}
+\sqrt{m_s m_b}m_b \lambda_{bs}\sin{\alpha}\cos{\alpha} \right ],
\label{f2hb0}
\eeq
where $x_i=m_i^2/\mhb^2$.

For the CP odd pseudoscalar Higgs boson $A^0$, we have
\beq
F_1^{A^0}&=&\frac{1}{4\mha^2} \sum_{i=s,b}
\left [ 3C_{11}(x_i) -C_1(x_i) \right ]
\sqrt{m_sm_b}m_i \lambda_{is}\lambda_{ib},
\label{f1a0}\\
F_2^{A^0}&=&\frac{1}{4 \mha^2} \sum_{i=s,b}
\left [ -m_b \left ( 3C_{11}(x_i) -C_1(x_i)\right ]\right. \nonumber\\
&& \left.
- 2m_i C_1(x_i) \right ] \sqrt{m_sm_b}m_i \lambda_{is}\lambda_{ib},
\label{f2a0}
\eeq
where $x_i=m_i^2/\mha^2$, and the functions $C_1(x_i)$ and
$C_{11}(x_i)$ in Eqs.(\ref{f1h0}-\ref{f2a0}) have been given in
Eqs.(\ref{c1xi},\ref{c11xi}).

In the numerical calculations, we  fix the following parameters
and use them as the standard input (SIP)
\cite{pdg98,buras96}: $M_W=80.41GeV$, $\alpha_{em}=1/129$,
$\sin^2{\theta_W} =0.23$,
$m_u =5$ MeV, $m_d =9$ MeV, $m_c =1.4$GeV,
$m_s=0.13$ GeV, $m_b=4.4$GeV, $ m_t=170$GeV,
$\Lambda^{(5)}_{\overline{MS}}=0.225$GeV,
$A=0.84$, $\lambda=0.22$, $\rho=0$, and $\eta=0.36$.
For the definitions and values of these input parameters, one can
see refs.\cite{pdg98,buras96}. For the running of $\alpha_s(\mu)$,
we use the two-loop formulae as given in ref.\cite{buras96} and
take $\mu =m_b$.

Using the SIP and assuming $m_{Higgs} =300GeV$ and $q^2=(1 - 19)
GeV^2$, we find numerically that:

\begin{itemize}
\item
In the SM, $|F_1^L|\approx 0.25$, $|F_2^{L,R}|\approx 0.008$;
For all neutral Higgs bosons, $F_1$ and $F_2$ form factors are
less than $10^{-4}$ in magnitude and therefore can be neglected
safely;

\item
For charged Higgs boson, $\tilde{F}_1^{L,R}$ are very small
and therefore can be neglected, but $\tilde{F}_2^L \approx 1.2 $
is larger than its
SM counterpart by about two orders of magnitude. While the
$\tilde{F}_2^R \approx 0.009$ is comparable with $F_2^R$ in the SM.
\end{itemize}

In the following, we will only consider the new effects of
$\tilde{F}_2^{L,R}$ form factors induced by charged Higgs boson.
The $q^2$ dependence of $F_1$ and $F_2$ in the SM and the
$\tilde{F}_2^{L,R}$ in the Model III comes from the penguin diagrams
with $c$ internal quarks. But this dependence is very
weak and therefore can be neglected safely.


Now we turn to calculate the
inclusive charmless decay $b\to s g$ for both on-shell
and off-shell ($q^2 > 0$) gluon, in order  to check the possible
enhancements to the decay widths and branching ratios induced by
the charged Higgs boson in Model III.

For the decay $b \to sg $ with an on-shell gluon as depicted in
Fig.2a, it is easy to derive the decay rate
\beq
\Gamma(b \to sg)^{SM}=\frac{2 \alpha_s(m_b)}{\pi}
\Gamma_0 \; |F_2|^2 \label{gammasm}
\eeq
in the SM, and
\beq
\Gamma(b \to sg)^{III}
=\frac{2 \alpha_s(m_b)}{\pi}\Gamma_0 \left ( |\bar{F}_2^R|^2 +
\frac{m_s^2}{m_b^2}|\bar{F}_2^L|^2 \right )\label{gamma3}
\eeq
in the Model III.
Here $\Gamma_0=G_F^2m_b^5/(192\pi^3)$, $ \bar{F}_2^{L,R} =
F_2 + \tilde{F}_2^{L,R}$, the SM form factor $F_2$ has been given in
Eq.(\ref{f2lr0}). For $\mhp = 200$ GeV, one finds
\beq
R_g&=&\frac{\Gamma(b \to sg)^{III}}{\Gamma(b \to sg)^{SM}}
=14.6 \label{rga}
\eeq
and the mass dependence of $R_g$ is shown in Fig.3a

The branching ratio $\brbsg$ is of the form
\beq
\brbsg= \frac{3\alpha_s(m_b) \cdot 0.105}{ \pi
|V_{cb}|^2 f(z) \kappa (z)} \left \{
\begin{array}{l}|F_2|^2, \ \ in  \ \ SM, \\ \left(
|\bar{F}_2^R|^2 + \frac{m_s^2}{m_b^2}|\bar{F}_2^L|^2,\right )
 \ \ in  \ \ Model \ \ III, \\
\end{array} \right. \label{brbsg}
\eeq
where $f(z)= 0.54$ is the phase space factor in the
semileptonic b-decay with $z=m_c/m_b$, $\kappa(z)= 0.88$
is the QCD factor to the semileptonic b-decay, and
$BR(B \to X_c e\bar{\nu}_e) =10.5\%$\cite{pdg98}.

Figs.(3a) and (3b) are the plots of  $R_g$ and the ratio
$\brbsg$ versus $\mhp$ in all three types of 2HDM for $\mhp =(100-800)$ GeV.
For theoretical predictions in Model I and Model II, we directly used
the formulae given in ref.\cite{hou94}.
For $\mhp =200$GeV, the branching ratio $\brbsg$ will be increased
from $\sim 0.2\%$ in the SM to $\sim 2.6\%$ in the Model III, as shown by
the upper solid curve in Fig.3b. The short-dashed and dotted curve in
figs.(3a) and (3b) shows the ratios in Model I and Model II, respectively.
It is easy to see that the enhancements to the ratios in both Model I and
II  are much smaller than that in Model III.

For $q^2 > 0$, the gluon is virtual and time-like, therefore before
it can fragment into real hadrons, it first disintegrates into real
partons such as on-shell quark pair $q \bar{q}$ for $q=(u,d,s)$
and gluon pair $gg$, as shown in Fig.2. The process $b \to u\bar{u}$
and $b \to d \bar{d}$ can be treated on the same footing, while one should
take into account the identical particle effects for the decay
$b \to s s \bar{s}$. Here, as an illustration, we will evaluate the
new contributions to the branching ratio
$\brbsqq$ for $q=u,d,s$ in the Model III. We use the same method
as ref.\cite{hou88}.

In order to illustrate the effects of charge Higgs boson on the
differential decay rate, we draw the Fig.4 for the typical decay mode
$\bsuu$.
\beq
\frac{d\Gamma(\bsuu)}{dy} &=&\frac{\alpha_s^2(m_b)
}{6\pi^2} \Gamma_0 \left \{
(1-y)^2(1+2y) |F_1^L|^2 \right. \nonumber\\
&&\left. +\frac{1}{y} [ 2 - 3y +y^3 ]\left [
 |\bar{F}_2^R|^2
 + \frac{m_s^2}{m_b^2}|\bar{F}_2^L|^2 \right ]
-6 (1-y)^2 Re\left [ (F_1^L)^*\bar{F}_2^R
 \right ] \right. \nonumber\\
&&\left. + \frac{6 m_s^2}{m_b^2}(1-y^2)
 Re\left [ (F_1^L)^*\bar{F}_2^L \right ]
- \frac{12 m_s^2}{m_b^2} (1-y)  Re \left [
 (\bar{F}_2^R)^*\bar{F}_2^L \right ] \right \},  \label{dgdy}
\eeq
where $y=q^2/m_b^2\approx 1-2E_s/m_b$. If one treats the $E_s$ as the
kaon energy, then the differential rate in Eq.(\ref{dgdy}) can be
regarded as the "Kaon-energy" spectrum. In Fig.4, the $c\bar{c}$
threshold cusp is clearly exhibited, where the short-dashed (solid)
curve shows the differential rate in the SM (Model III). It is easy
to see that the new contribution is peaked at the lower $q^2$ region.

The decay width $\Gamma(\bsqq)$ with $q=(u,d,s)$ is of the form
\beq
\Gamma(\bsqq)&=&\frac{\alpha_s^2(m_b)
}{12\pi^2} \Gamma_0 \left \{ \frac{17}{6}|F_1^L|^2
 -\frac{34}{3} Re\left [ (F_1^L)^*\bar{F}_2^R \right ] \right.
\nonumber\\
&&\left. -  \left [ \frac{49}{3} +10 \log[y_{min}] + 2\log[x_{min}] \right]
\left [  |\bar{F}_2^R|^2 + \frac{m_s^2}{m_b^2}|\bar{F}_2^L|^2
 \right ] \right.\nonumber\\
&&\left. + \frac{24 m_s^2}{m_b^2} Re\left [ (F_1^L)^*\bar{F}_2^L\right]
- \frac{30 m_s^2}{m_b^2} Re \left [
 (\bar{F}_2^R)^*\bar{F}_2^L\right ] \right \},  \label{bsqq}
\eeq
where $x_{min}=y_{min}=4m_s^2/m_b^2$. Contrary to the case of decay
$b \to sg$ with on-shell gluon, the new contribution from charged Higgs
boson in Model III tend to decrease the decay width $\Gamma(b \to s q
\bar{q})$ slightly by  about five percent with respect to the SM
prediction. The reason for this behaviour is simple. We know that only
the new magnetic $\tilde{F}_2$ form factor  contribute effectively
to the decay processes under study. For the decay $b \to sg$
with on-shell gluon, the magnetic form factor $\tilde{F}_2$ dominate
the total contribution. But for the case of $q^2 >0$, the SM form factor
$F_1^L$ control the decay processes, the new $\tilde{F}_2$ can
contribute only through interference with the SM $F_1^L$.
The large infrared(IR)
logarithms in Eq.(\ref{bsqq}) will be canceled if one make a
complete $O(\alpha_s)$ treatment of the second term in
Eq.(\ref{bsqq}). Since we
are not giving a full $O(\alpha_s)$ QCD analysis, we can not
include this term consistently. We therefore simply drop it from
further discussions, as done in ref.\cite{hou88}.

The decay mode $b \to s gg$ {\em via or not via} $g^* \to gg$
has been studied in ref.\cite{hou88}. It was found that the decay
width $\Gamma(b \to s gg)$ depend on the form factor $F_1$ only,
\beq
\Gamma(b \to s gg)=\frac{\alpha_s^2(m_b)}{4 \pi^2} \Gamma_0
|F_1^L|^2\label{bsgg}
\eeq
As discussed previously, the new contributions to the electric form
factor $F_1$ from charged and neutral Higgs bosons are very small and
has been neglected. The decay $b \to s g g$  therefore will be
not affected effectively by gluon penguins induced by
Higgs bosons appeared in Model III.

Collectively, the branching ratio $BR(b \to s g^*)$ (here $b \to sg^*$ is
symbolic for processes of $b \to sg$ (on-shell gluon), $b \to s q \bar{q}$,
and $b \to s gg$ ) can be written in the form
\beq
BR(b \to s g^*)^{SM}&=& \frac{0.105}{|V_{cb}|^2 f(z) \kappa (z)}\left \{
\frac{3\alpha_s(m_b)}{\pi} |F_{2}|^2 \right. \nonumber \\
&&+\left.  \frac{\alpha_s^2(m_b)}{16\pi^2} \left [
\frac{35}{6}|F_1^L|^2
-\frac{34}{3} Re\left [ (F_1^L)^* F_2 \right ] \right ]\right \}
\label{brbsgssm}, \\
BR(b \to s g^*)^{III}&=& \frac{0.105}{|V_{cb}|^2 f(z) \kappa (z)}\left \{
\frac{3\alpha_s(m_b)}{\pi} \left [ |F_2|^2 +
\frac{m_s^2}{m_b^2}|\bar{F}_2^R|^2 \right ] \right. \nonumber\\
&&+\left.  \frac{\alpha_s^2(m_b)}{16\pi^2} \left [
\frac{35}{6}|F_1^L|^2
-\frac{34}{3} Re\left [ (F_1^L)^* \bar{F}_2^R \right ] \right.
\right. \nonumber\\
&&\left. \left.
+ \frac{24 m_s^2}{m_b^2} Re\left [ (F_1^L)^*\bar{F}_2^L\right]
- \frac{30 m_s^2}{m_b^2} Re \left [
 (\bar{F}_2^R)^*\bar{F}_2^L\right ] \right ]\right \}
\label{brbsgsnp},
\eeq

As illustrated in Fig.5, the branching ratio $\brbsgs = 5.5\% - 1.6\%$ for
$\mhp =100 - 800$ GeV in the Model III (solid curve) compared with
$1.3\%$ in the SM (short-dashed line). For light charged Higgs,
the enhancement is significant: $\brbsgs = 3.7\%$ for
$\mhp =200$ GeV in the Model III. The upper dot-dashed line in
Fig.5 shows the CLEO upper limit: $BR(b \to s g^*) < 6.8\%$ at $90\%
C.L.$\cite{cleo98b}.  In ref.\cite{wu99}, the author studied
the CP violation in radiative B decays in a type-III 2HDM with
the fourth generation fermions, and found that the branching ratio
$\brbsg $ can be increased to $10\%$ level.

To summarize, we have calculated, from the first principle,
the new gluon penguin diagrams that contribute to the inclusive
charmless decays of b quark in the Model III without inclusion of
QCD corrections. We start from the
evaluation of the new gluon penguin diagrams induced by the
exchange of charged and neutral Higgs bosons $(H^{\pm}, h^0,
H^0$, and $A^0)$,  derive out the
$F_1$ and $F_2$ form factors which control the new contributions to
the  inclusive b quark decays under study, and finally calculate the
relevant decay rates and branching ratios. We found that:

\begin{quotation}

(a) Among charged and neutral Higgs bosons, the CP even
charged Higgs boson $H^{\pm}$ dominate the new contribution. By using
the assumed FC couplings, the contribution of the
neutral scalar and pseudo-scalar is completely negligible. Therefore,
both the value of the mixing angle $\alpha$ and masses $(m_{h^0},
m_{H^0}$, and $m_{A^0} )$ are irrelevant.

(b) The new electric form factor $\tilde{F}_1$ is also completely
negligible. The new magnetic form factor $\tilde{F}_2^L$, on the
contrary,  is much larger than its SM counterpart and thus contribute
significantly to the decay processes in question. The $\tilde{F}_2^R$
is comparable in size  with the $F_2$ in the SM and can also
contribute effectively.

(c) The charged Higgs enhancement to the branching ratio $\brbsg$
with $q^2=0$ can be as large as a factor of $25$ ($14.6$) for
$\mhp=100 GeV$ ($200 GeV$) in the Model III.
So large enhancement will be very
helpful to generate a large ratio $\brbsg$ favored by some
experimental data such as the deficit in charm counting,
a $3\sigma$ deficit in kaon counting as well as the well-known large
$B \to \eta' X_s$ branching ratio measured by the CLEO collaboration.
The enhancement in the Model III can be much larger than
that in both Model I and II.

(d) For the subprocesses with a off-shell gluon ($q^2 > 0$), the new
contribution to the decay mode $b \to s q \bar{q}$ with $q=(u,d,s)$
is numerically small and peaked at the lower $q^2$ region. And the new
contribution to $b \to s gg$ can be neglected.

\end{quotation}

\vspace{1cm}
\hspace*{4.5cm}ACKNOWLEDGMENTS\\

This work was supported in part by the National Natural Science
Foundation of China, a grant from the State Commission of Science
and technology of China and the Doctoral Program
Foundation of Institution of Higher Education. Z.J. Xiao
acknowledges the support by the National Natural Science Foundation
of China under the Grant No.19575015, and by the funds from the
Outstanding Young Teacher Foundation of the Education Ministry of
China and the funds from Science and Technology Committee of Henan
Province.

\newpage
\begin{center}
{\bf Figure Captions}
\end{center}
\begin{description}

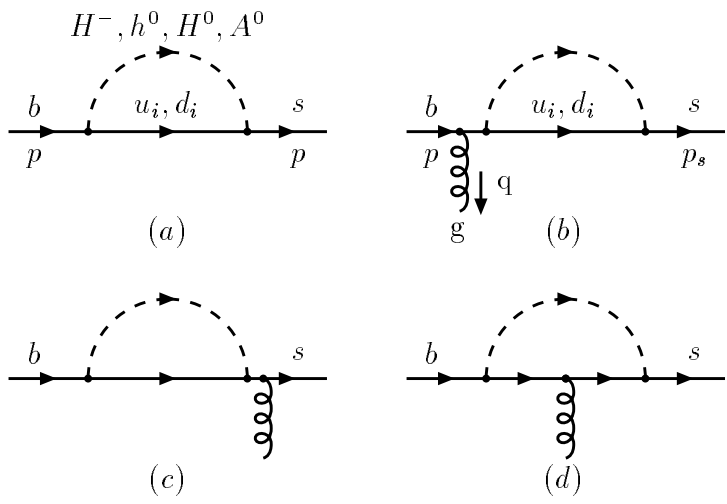
\item[Fig.1:] Self-energy and gluon penguin diagrams with charged and neutral
scalar or pseudoscalar exchanges in the Model III. The charged  and
neutral Higgs boson propagators correspond to the up-type quarks
$u_i=(u,c,t)$ and the down-type quarks $d_i =(d,s,b)$, respectively.

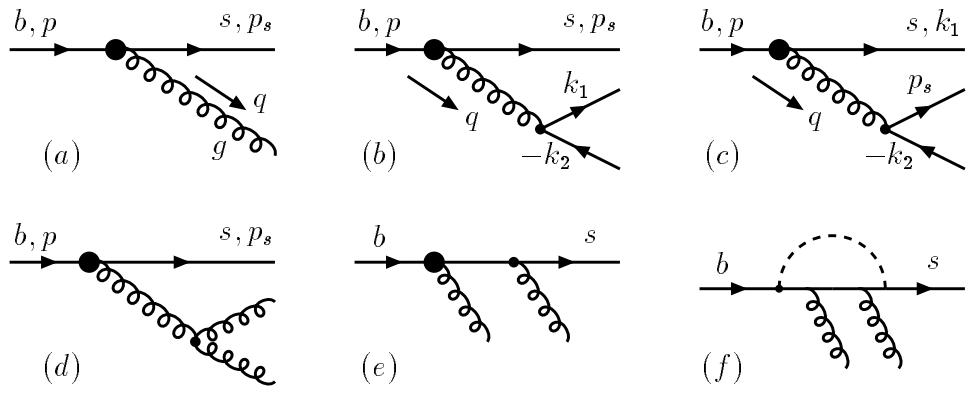
\item[Fig.2:] Feynman diagrams for the charmless decay of b quark in the
Model III: (a) $b \to s g$ ($q^2=0$, on-shell gluon); (b) $b \to s
q\bar{q}$ ($q^2 > 0$) with $q=u,d,s$; (c) additional diagram for
$b \to s s\bar{s}$;(d)  $b \to s g g$ {\em via} $g^* \to gg$;
(e) typical  one-particle reducible diagram that lead to
$\bsgg$; (f) typical one-particle irreducible diagram that lead to
$\bsgg$. The blobs in first five diagrams  denote the effective $bsg$
vertex.

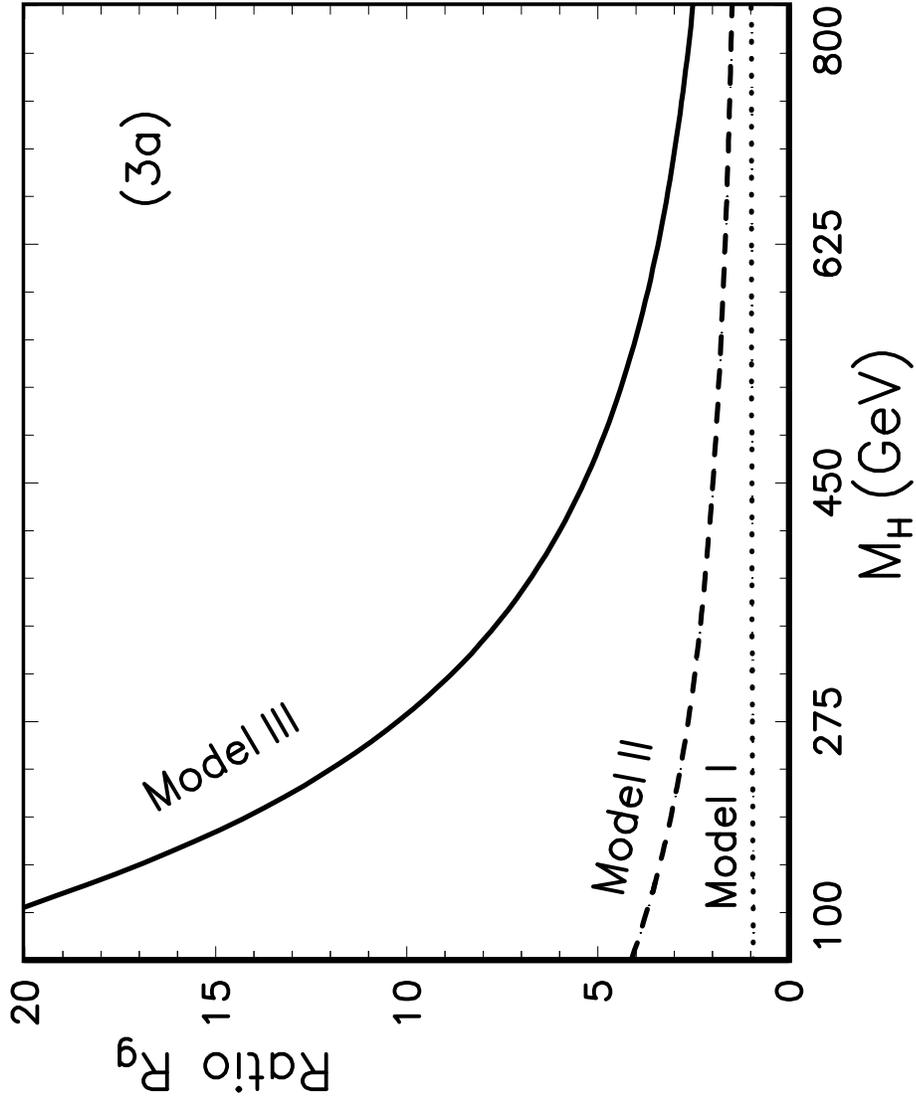
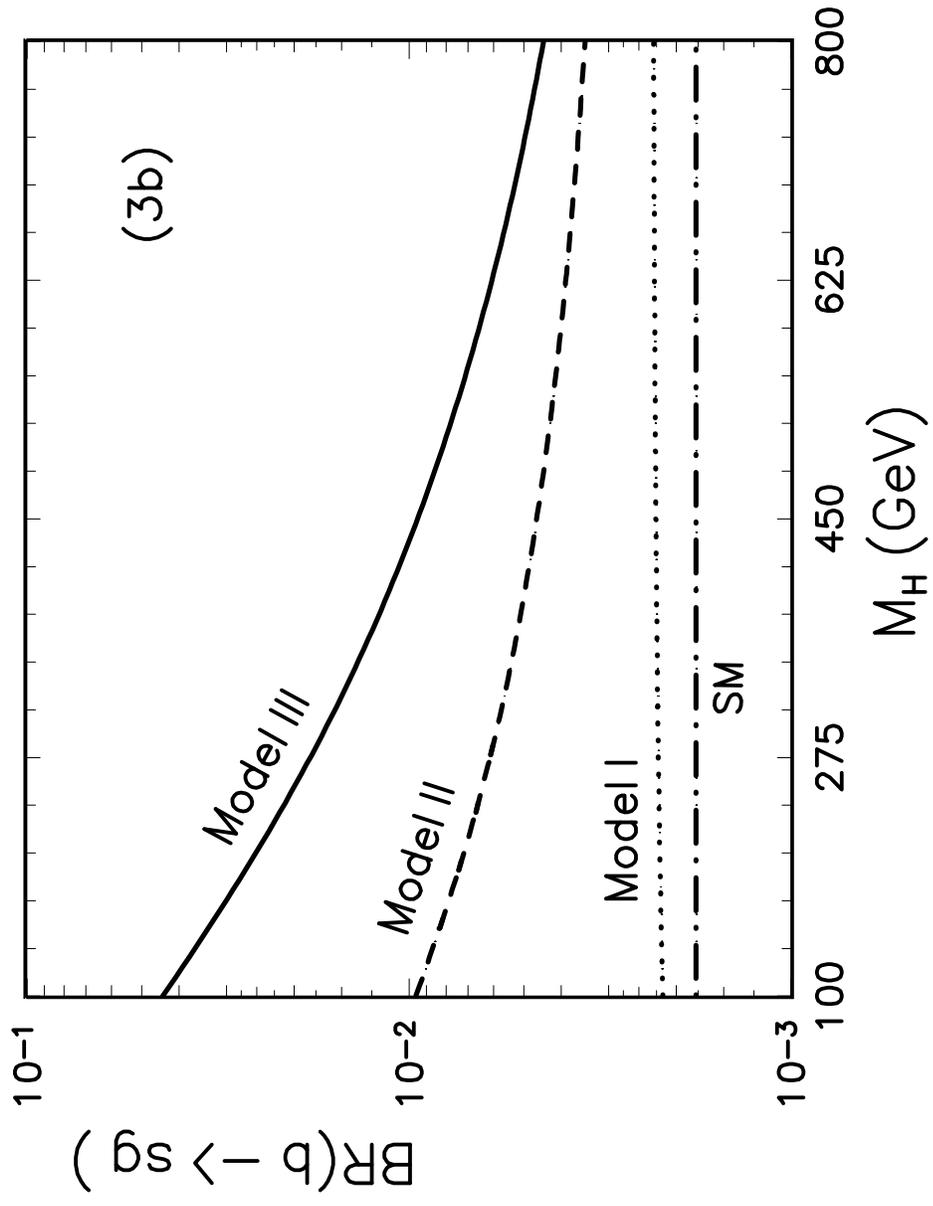
\item[Fig.3:] Mass dependence of the ratios $R_g$ and $\brbsg$ ($q^2=0$) in
three type of two-Higgs-doublet models for $\mhp=100-800 GeV$.
In Fig.3a and (3b), the dotted, short-dashed and solid curve shows the ratios
in the Model I, II and III, respectively. The dot-dashed line in
Fig.3b is the SM prediction.

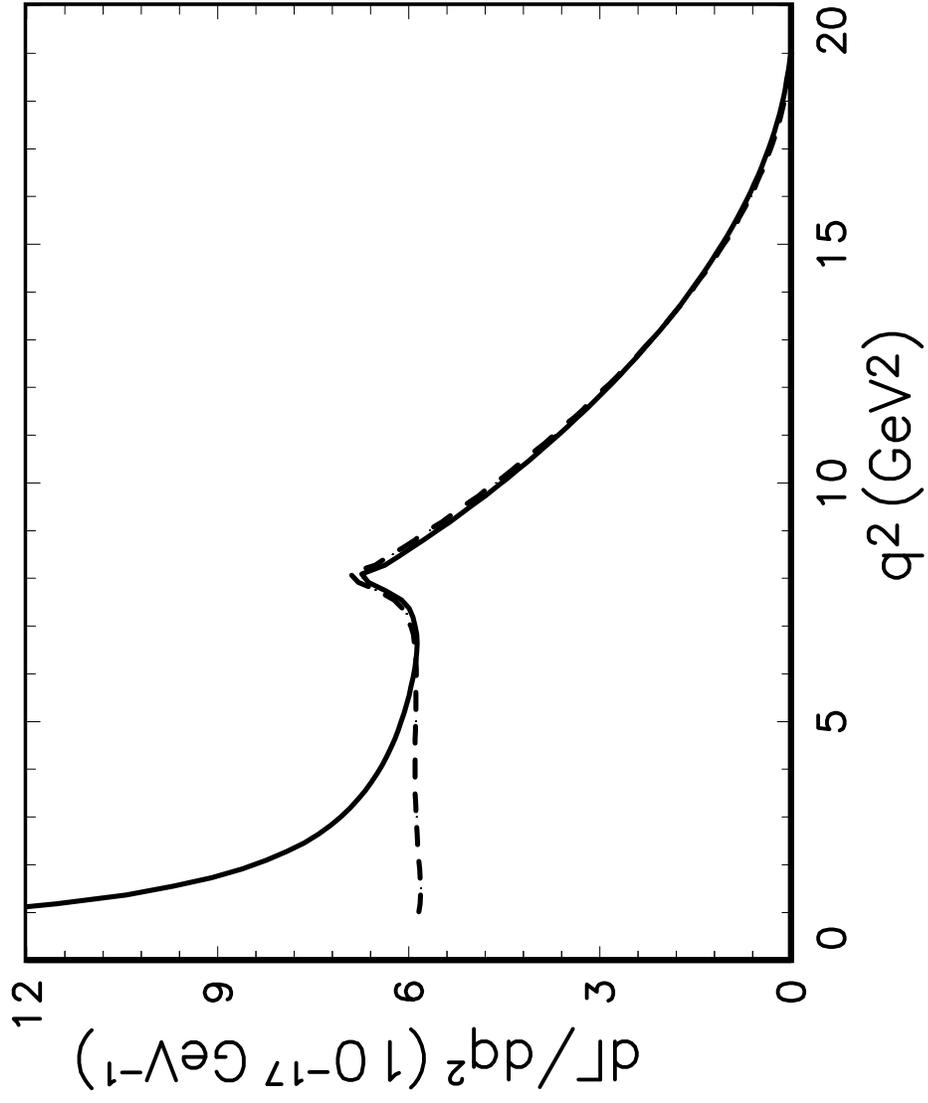
\item[Fig.4:] The $q^2$ dependence of differential decay
rates for $b \to s u \bar{u}$. The short-dashed  and
solid line shows the rate in the SM and Model III, respectively.

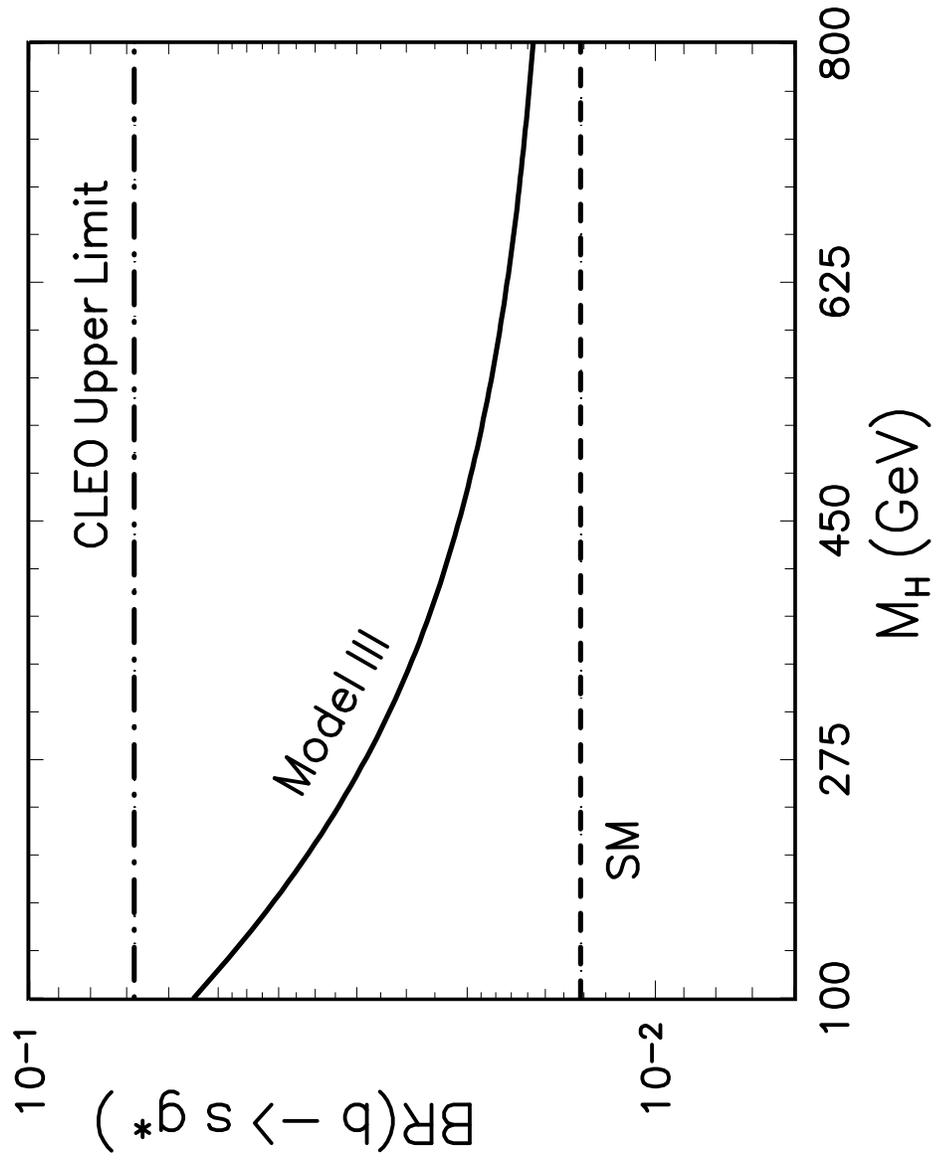
\item[Fig.5:] The mass dependence of branching ratio $\brbsgs$ for $\mhp=
100-800 GeV$. The short-dashed and solid curve shows the ratio in the SM
and the Model III, respectively.

\end{description}

\end{document}